\documentclass[lettersize]{IEEEtran}
\usepackage{amsmath,amssymb,amsfonts}
\usepackage{array,booktabs,multirow}
\usepackage[most]{tcolorbox}
\usepackage{algorithm,algorithmic}
\usepackage{graphicx}
\usepackage{url}
\usepackage{cite}
\usepackage[T1]{fontenc}
\usepackage{newtxmath}

\newtcolorbox{prompttemplate}[1]{
  breakable,
  sharp corners,
  colback=black!2,
  colframe=black!40,
  colbacktitle=black!10,
  coltitle=black,
  fonttitle=\bfseries\footnotesize,
  title={#1},
  boxrule=0.4pt,
  left=6pt,right=6pt,top=5pt,bottom=5pt,
  before skip=6pt,after skip=6pt
}

\begin{document}
\title{SatCraft: Agentic Optimization of Scheduling Policies for Satellite--Cellular Edge Services}
\author{Pengcheng Luo, Dengke Deng, Zenghui Zhang, Genke Yang, and Jian Chu%
\thanks{This work was supported by the National Science and Technology Major Project of China under Grant 2025ZD1602400. \emph{(Corresponding authors: Zenghui Zhang and Genke Yang.)}}%
\thanks{Pengcheng Luo, Dengke Deng, Zenghui Zhang, Genke Yang, and Jian Chu are with Ningbo Artificial Intelligence Institute, Shanghai Jiao Tong University, Ningbo 315000, China, and also with the School of Automation and Intelligent Sensing, Shanghai Jiao Tong University, Shanghai 200240, China, and the Key Laboratory of System Control and Information Processing, Ministry of Education of China, Shanghai 200240, China (e-mail: luopeng69131@sjtu.edu.cn; pokier123@sjtu.edu.cn; zenghui.zhang@sjtu.edu.cn; gkyang@sjtu.edu.cn; chujian@niii.com).}%
\thanks{The source code will be made available at \protect\url{https://github.com/luopeng69131/SatCraft}.}%
}

\maketitle
\thispagestyle{plain}

\begin{abstract}
Mobile edge services can combine local computation with satellite and cellular access to deliver time-sensitive results in areas with intermittent terrestrial coverage. Meeting their deadlines and quality requirements calls for joint computation and transmission scheduling that accounts for task dependencies, changing link capacity, and shared resource budgets. However, optimizing resource allocations or tuning parameters within a fixed policy structure offers limited scope to correct failures rooted in the scheduling logic. We propose \emph{SatCraft}, a framework in which a large language model (LLM) agent uses execution feedback to revise the code of joint scheduling policies. Each revision undergoes policy checks and complete-run evaluation to assess its validity, runtime, and effects on service completion. Because policy performance varies with access conditions, SatCraft retains valid alternatives in a policy library and selects one for each episode using initial access observations and the known resource configuration. Experiments using Starlink and cellular measurement data show that SatCraft improves cumulative completed-service value by 3.61\%--16.89\% over baselines, including recent state-of-the-art scheduling methods ADPRL, EdgeOPT, and TF-DDRL. It achieves the highest completed-service value in all four resource profiles and averages 25.02~ms per scheduling call. These results demonstrate how LLM-guided policy optimization can improve complete service delivery under the computing, communication, and quality constraints of satellite--cellular networks.
\end{abstract}


\section{Introduction}
\subsection{Background and Motivation}
\IEEEPARstart{S}{atellite} and cellular connectivity extend the reach of mobile edge services in areas with intermittent terrestrial coverage. Timely delivery requires coordinating computation and communication under shared resource constraints~\cite{du2025online,yao2026dualdependency}. In field inspection, a vehicle equipped with cameras and an onboard computer continually collects observations for a remote assessment team. When a road anomaly is detected, the team may need both an urgent report and usable image evidence before a deadline. Routine image uploads continue to arrive, competing for the same computing and communication resources. The vehicle can send original observations or process them locally to produce smaller outputs, with satellite and cellular links providing complementary upload paths.

\begin{figure}[!t]
  \centering
  \includegraphics[width=\linewidth]{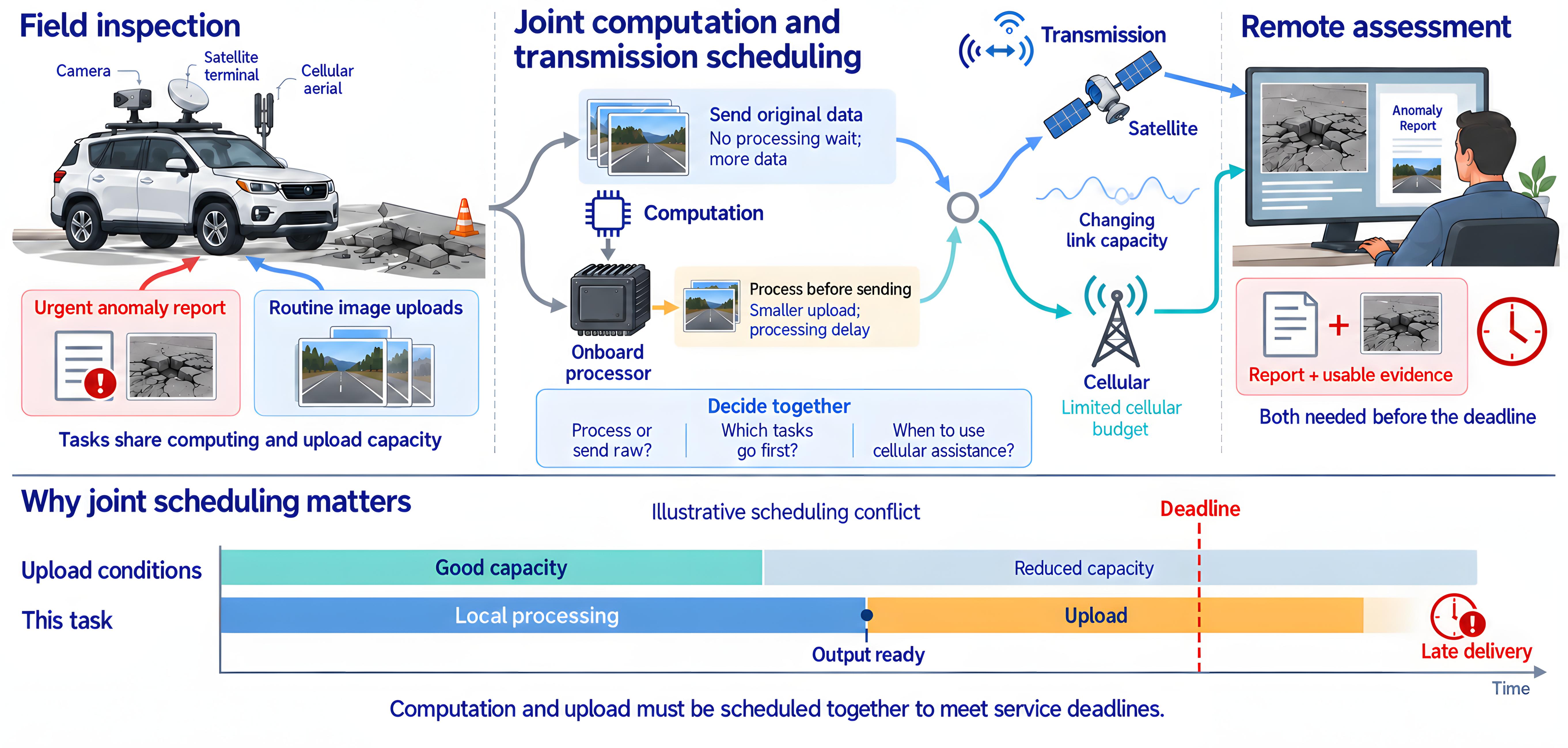}
  \caption{Field inspection over satellite--cellular access and an illustrative scheduling conflict.}
  \label{fig:intro-background}
\end{figure}

These choices create a tradeoff between computing delay and transmission demand~\cite{yang2024compression,chen2025slice}. Sending original data avoids local processing delay but requires more transmission capacity; processing first produces a smaller output that becomes ready later. If processing outlasts a period of good link conditions, even a smaller upload may miss its deadline as capacity declines. Competition from other tasks and dependencies between processing stages can introduce further delays~\cite{zhang2024dependency}, while cellular assistance consumes a limited usage budget. An assessment may remain incomplete even when its report arrives on time if the required evidence is missing or unusable. Meeting these requirements calls for joint computation and transmission scheduling that coordinates processing choices, task priorities, and cellular use to deliver complete, acceptable services on time. Fig.~\ref{fig:intro-background} illustrates this setting and a possible timing conflict.

Recent research provides important foundations for such joint decisions. Fan \emph{et al.} optimize inference offloading and resource allocation in satellite--ground networks~\cite{fan2025satellite}; SLICE coordinates layer-wise computation across satellite and ground resources~\cite{chen2025slice}; and Shisher \emph{et al.} co-schedule feature generation and transmission for timely multi-task inference~\cite{shisher2025coscheduling}. These approaches establish the value of coordinating computation and communication. For a continuing stream of services, however, a scheduling policy must also translate task progress, result dependencies, and remaining budgets into decisions that remain effective as access conditions change. Execution traces can reveal why an urgent service failed: a processing choice delayed its output, a priority rule postponed required evidence, or premature cellular activation depleted its allowance. Improving such behavior requires revisiting the decision logic and assessing its consequences throughout service execution. Constructing and adapting executable joint scheduling policies is thus a practical challenge alongside resource optimization.

Large language model (LLM) agents offer a way to connect such execution failures to changes in scheduling logic. By interpreting service requirements, policy code, and execution feedback together, an agent can propose changes to the conditions and priorities that govern computation and transmission. Work on program search, reflective heuristic evolution, and tool-assisted scheduler development demonstrates the potential of this approach~\cite{romeraparedes2024funsearch,ye2024reevo,zheng2025schedcp}. For satellite--cellular services, failures such as delayed evidence or premature cellular activation can guide revisions that account for output readiness, required products, and remaining budgets. These revisions must be assessed through complete service executions: helping one task can consume resources needed by later arrivals, so an apparently useful local change may reduce overall delivery. Access variability adds a further consideration, since scheduling logic that is effective under one access pattern may be less useful under another. This suggests improving joint scheduling logic through complete-service feedback and retaining alternative policies for selection based on initial access observations and known resource conditions.

\begin{figure}[!t]
  \centering
  \includegraphics[width=\linewidth]{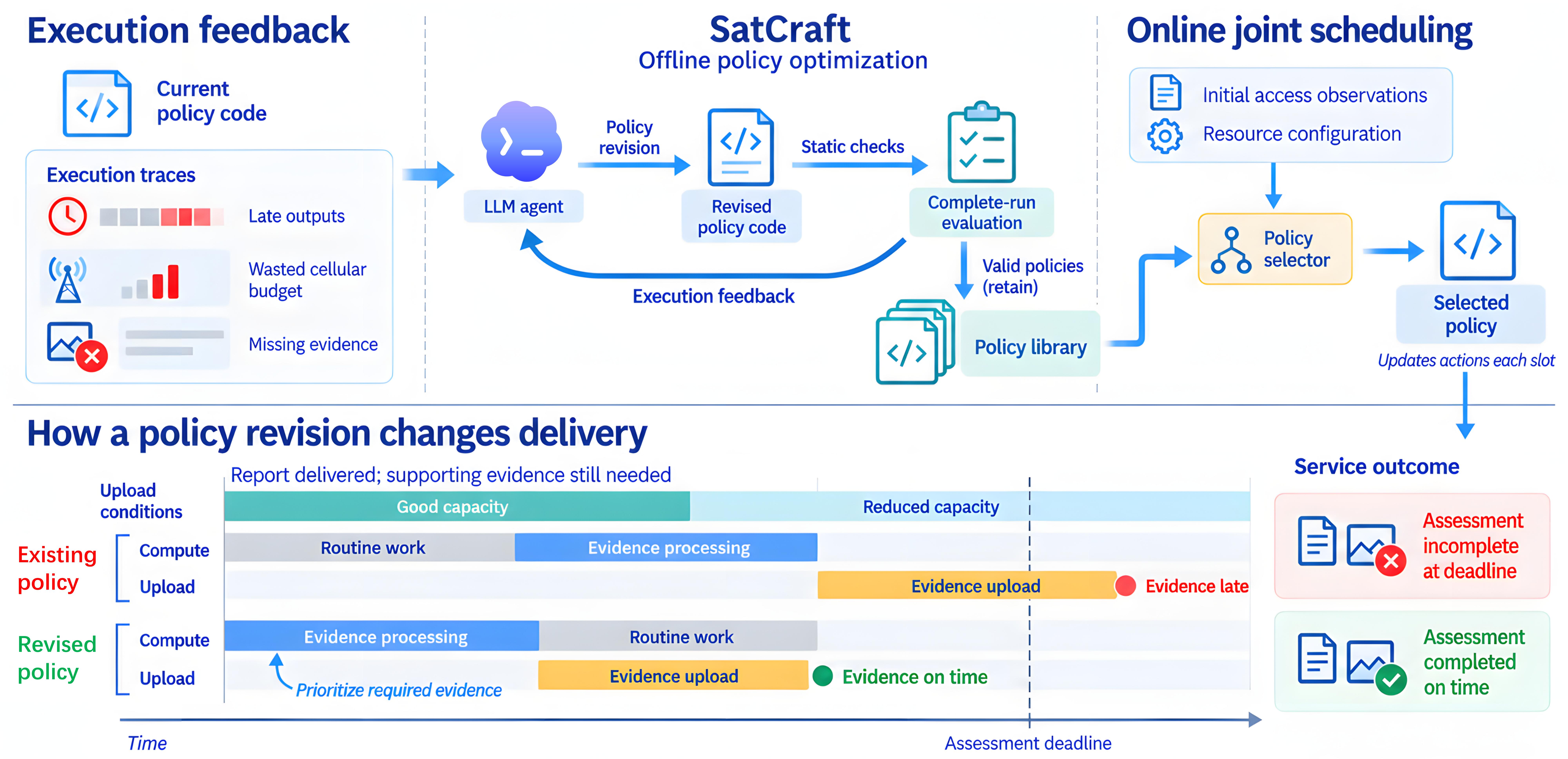}
  \caption{SatCraft policy optimization and its effects on joint scheduling and service completion.}
  \label{fig:intro-satcraft-concept}
\end{figure}

We develop \emph{SatCraft}, an agentic framework for optimizing scheduling policies for satellite--cellular edge services. An LLM agent constructs and repairs joint scheduling policies using feedback from complete service executions. Policy checks and complete-run evaluation assess validity, runtime, and completed-service value; a policy library retains valid alternatives. SatCraft selects one policy from initial access observations and known resource conditions. During execution, that policy jointly determines processing modes, computing and transmission priorities, and cellular use as tasks progress and quality becomes known. Fig.~\ref{fig:intro-satcraft-concept} connects this policy optimization process to changes in task schedules and their service outcomes. SatCraft thereby turns execution feedback into reusable scheduling logic that coordinates scarce computing and access resources around timely, acceptable service completion.

\subsection{Contributions}
The main contributions of this paper are as follows:
\begin{itemize}
\item We formulate a dual-access scheduling model around complete service delivery. It jointly represents output timing after local processing, delivered-result quality, task dependencies, and satellite and cellular resource limits. The objective is the cumulative value of services whose acceptance conditions are met before their deadlines, allowing processing and transmission decisions to be assessed by their downstream effects.

\item We develop an execution-feedback-driven method for optimizing scheduling policies. SatCraft uses an LLM agent to revise policy code according to specified repair directions and execution feedback. Interface checks and complete-run evaluation admit policies that satisfy validity and runtime constraints to a policy library, where changes to load anticipation, link activation, and quality-dependent paths can be evaluated and reused.

\item We design a mechanism for context-dependent policy selection and execution. A shallow policy selector relates development-set utility to initial dual-access observations and known resource configurations. After the initial access observation period, SatCraft selects one policy, which then adapts its actions to arriving tasks, revealed quality, and remaining resources.

\item We evaluate SatCraft using measured Starlink and cellular rates and modeled service workloads. SatCraft achieves the highest cumulative completed-service value among the compared methods, including ADPRL, EdgeOPT, and TF-DDRL. It also achieves the highest completed-service value across all four resource profiles, with gains of 3.80\%--30.16\% over these three methods under different resource pressures.
\end{itemize}

\subsection{Organization}
Section~II reviews satellite--terrestrial edge scheduling and LLM-driven policy construction. Section~III defines the dual-access service model and scheduling objective. Section~IV describes SatCraft's policy optimization, policy selection, and online joint scheduling. Section~V presents the experimental setting and evaluates overall performance, resource pressure, method components, optimization and runtime costs, and repair mechanisms. Section~VI concludes the paper.

\section{Related Work}
We review how prior work connects resource constraints to useful edge-service results, and how LLMs turn task feedback into executable policies. These two lines inform the scheduling objective and the construction of policies that pursue it.

\subsection{Edge-Service Scheduling with Satellite and Terrestrial Access}
Computing and access conditions determine which edge-service results can be delivered. Laniewski \emph{et al.} characterize throughput, latency, loss, and power consumption during vehicle-mounted Starlink operation~\cite{laniewski2024road}, while Wang \emph{et al.} measure simultaneous Starlink and cellular access to examine performance diversity and multipath opportunities~\cite{wang2025digitaldivide}. Processing is itself constrained: for onboard satellite workloads, Orbital Edge Computing examines pipelines under orbital, energy, and downlink limits~\cite{denby2020orbital}, and Xing \emph{et al.} measure the thermal and power constraints of in-orbit COTS devices~\cite{xing2024satellitecots}. Across these distinct deployment settings, communication opportunities and computing limits motivate decisions about where and how much processing to perform before transmission.

Collaborative inference addresses this allocation problem by coordinating processing placement and communication. Fan \emph{et al.} combine deep reinforcement learning and numerical optimization to select offloading locations and allocate resources under latency, energy, and accuracy considerations~\cite{fan2025satellite}. SLICE uses layer-level energy measurements to schedule deep neural network (DNN) partitioning and satellite--ground execution under a deadline~\cite{chen2025slice}. TargetFuse+ combines satellite--ground object counting with model protection and bandwidth- and energy-aware data throttling~\cite{zhang2026targetfuse}, while Shisher \emph{et al.} co-schedule feature generation and transmission for timely multi-task inference at the wireless edge~\cite{shisher2025coscheduling}. These approaches link processing effort to transmission demand and delivery time. Assessing the benefit of these choices also requires relating their resource costs to the quality of the resulting output.

Application-level adaptation makes this quality--resource relationship explicit. AWStream profiles accuracy--bandwidth tradeoffs to adapt data quality and sending rate~\cite{zhang2018awstream}; OneAdapt estimates how input configuration affects inference accuracy to guide rapid adaptation~\cite{du2023oneadapt}. For computing resources, Ekya uses performance profiles to schedule inference and retraining on shared edge GPUs while balancing resource demand and accuracy~\cite{bhardwaj2022ekya}. When an output feeds subsequent stages, however, the value of such adaptation also depends on its effect on the remaining workflow: a useful intermediate result must still enable the application to finish its required work.

Workflow optimization provides a framework for following these downstream effects. Vulcan constructs, configures, and places live machine learning (ML) analytics pipelines across edge tiers, with online adaptation after deployment~\cite{zhang2024vulcan}. Decima represents job dependencies as graphs and learns scheduling policies to optimize completion time under competing workloads~\cite{mao2019decima}. These studies connect resource decisions to progress through an entire workflow. For satellite--cellular services, this connection also involves delivered quality: a result may determine which downstream products remain necessary, while the resources spent delivering it affect other services. Joint scheduling therefore calls for relating quality-dependent task progress and shared resource use to the timely completion of acceptable services.

\subsection{LLM-Driven Policy Construction and Optimization}
LLM-based methods offer ways to improve the decision logic linking task information to service outcomes. NetLLM adapts LLMs to networking decisions, including bitrate selection and cluster job scheduling~\cite{wu2024netllm}, while Reflexion uses task feedback to guide subsequent trials through verbal reflection~\cite{shinn2023reflexion}. Task information and feedback can also guide revisions to the executable code implementing a policy. FunSearch couples generated functions with execution-based evaluation~\cite{romeraparedes2024funsearch}; Evolution of Heuristics (EoH) evolves heuristic descriptions and code together, and ReEvo uses reflection to guide heuristic search~\cite{liu2024eoh,ye2024reevo}. Treating code as the optimization object makes proposed changes directly testable through execution, with the evaluator determining which changes count as improvements.

The evaluator consequently shapes both the search direction and the practical value of the resulting code. Eureka uses reinforcement-learning outcomes to guide reward-program revisions~\cite{ma2024eureka}, while AlphaEvolve applies evaluator-guided code evolution to scientific and systems algorithms, including data-center scheduling~\cite{novikov2025alphaevolve}. For systems scheduling, evaluation must also account for workload behavior and execution constraints. SchedCP integrates workload analysis, a scheduler repository, and validation interfaces for eBPF scheduling policies~\cite{zheng2025schedcp}. AI-Driven Research for Systems (ADRS) connects iterative program improvement to evaluators for cloud scheduling, load balancing, and transaction scheduling~\cite{cheng2025adrs}. Such evaluation supports measurable code improvement, but its results are tied to the workloads and constraints under which candidates are assessed. Reusing the resulting policies therefore also requires considering how their effectiveness varies across operating conditions.

This variation motivates retaining policies with complementary strengths and deciding when to use them. Multi-objective Evolution of Heuristic (MEoH) exposes performance--efficiency tradeoffs by evolving heuristics under multiple objectives~\cite{yao2025meoh}. Algorithm selection offers a way to exploit differences among candidates: SATzilla selects solvers using instance features and empirical difficulty models~\cite{xu2008satzilla}. Within LLM-based program search, EoH-S constructs complementary heuristics for different instance sizes and distributions~\cite{liu2026eohs}, while LACE uses problem interfaces, a tool library, and time-constrained search to discover complementary heuristics~\cite{gong2026lace}. These studies connect candidate diversity to differences in problem instances and resource demands, making the conditions under which a policy is useful part of the policy-optimization problem.

For satellite--cellular services, those conditions interact with dependencies that unfold over a complete execution. A policy revision can change when outputs become ready, whether delivered results enable downstream tasks, and how much capacity remains for later arrivals. Evaluation must follow these effects through the completion of acceptable services, while selection must associate policy performance with access observations and resource conditions available at deployment. Relating complete-service feedback to the context in which a policy will be used is therefore an important consideration when adapting LLM-driven policy optimization to joint computation and transmission scheduling.

\section{Dual-Access Edge-Service Model and Scheduling Objective}
Consider a mobile node with local computing resources and both Starlink and cellular access. An inspection vehicle, for example, must deliver an anomaly report and usable evidence to complete an urgent assessment. We use \emph{service} for an application-level deliverable and \emph{product task} for processing and transmission work that contributes to it. The report and evidence can be modeled as separate product tasks of one assessment service, sharing resources with routine uploads. More generally, a service may require several products or accept alternative combinations of them.

All products follow a common execution sequence. Once a product arrives and its dependencies are satisfied, the policy chooses either to transmit its raw data or to process it locally first. The resulting output enters a shared transmission queue and can be delivered over either access path. Quality is revealed upon delivery and can then enable downstream products or complete a service. Time is slotted: an action at slot $t$ executes over $[t,t+1)$, and a delivery completed in that slot is recorded at $t+1$. We first define service completion, then product execution and shared resources, and finally the policy's information, actions, and objective.

\begin{figure}[!t]
  \centering
  \includegraphics[width=\linewidth]{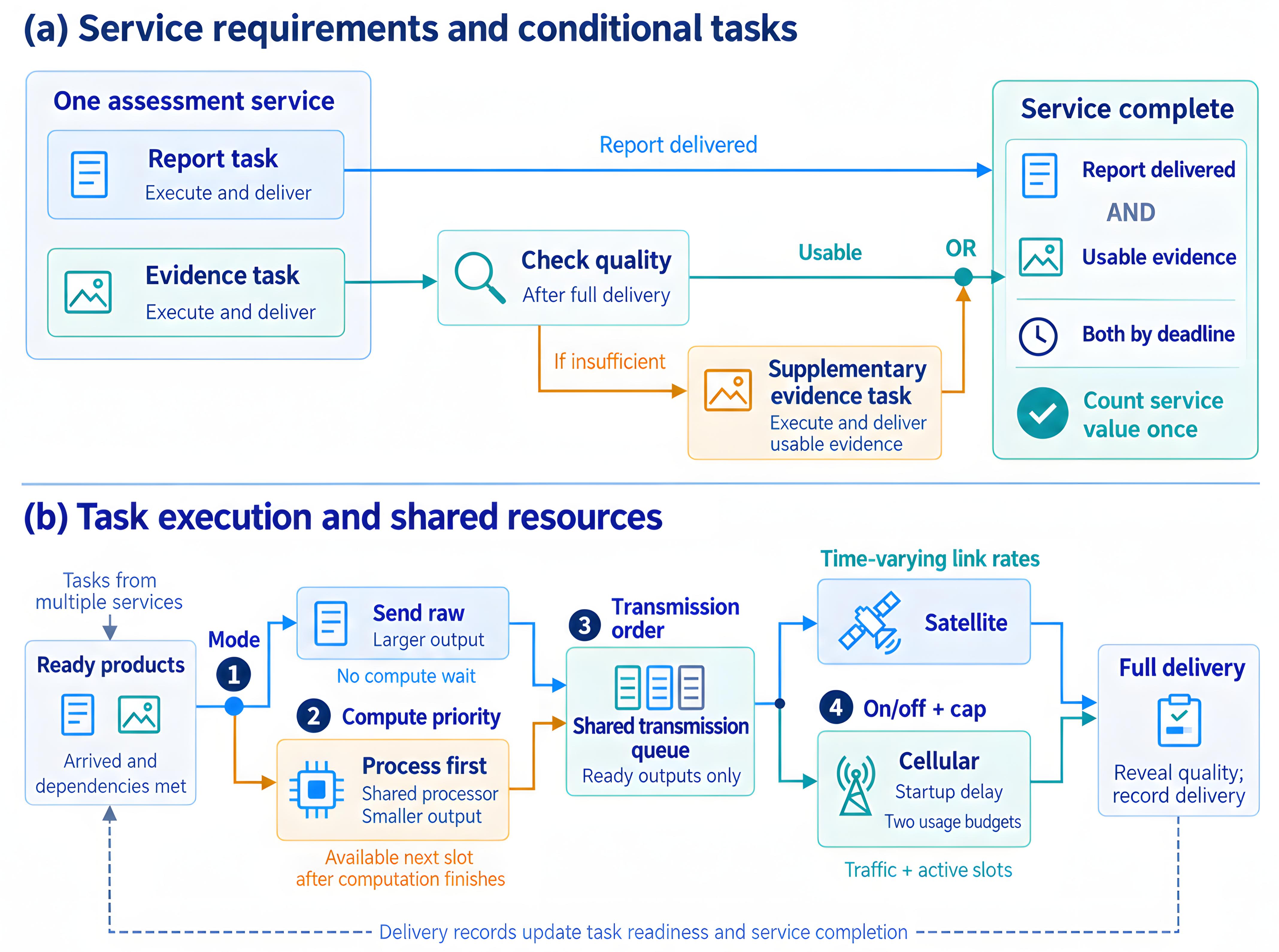}
  \caption{Service completion and joint computation--transmission scheduling.}
  \label{fig:problem-model}
\end{figure}

\subsection{Service Requests and Completion Conditions}
Delivered results affect both task readiness and service completion. Fig.~\ref{fig:problem-model}(a) illustrates these two roles: insufficient initial evidence quality enables a supplementary task, while completing the assessment requires a report and usable evidence by the deadline. We model these conditions using product-dependency and service-acceptance predicates evaluated on delivery records.

Let $\mathcal R_t$ be the set of product-delivery records available at time $t$. Each record contains the product identity, delivery time, processing mode, and revealed quality. Product $i$ has a dependency predicate $h_i(\mathcal R_t)$, formed by conjunctions or disjunctions of required deliveries and quality conditions. Only an arrived product whose predicate holds can choose a mode and begin execution; a product without predecessors becomes ready on arrival. Quality-dependent predicates also express confirmation or substitute paths, enabling the policy to determine which outstanding work remains useful given revealed results.

Service $s$ is specified by its arrival time $a_s$, deadline $d_s$, completion value $w_s$, and acceptance predicate $g_s(\mathcal R_t)$. The predicate similarly combines delivery and quality conditions. After each slot updates the delivery records, an active service is checked for its first qualifying completion time:
\begin{equation}
\tau_s=\inf\{t\in\{a_s+1,\ldots,d_s\}:g_s(\mathcal R_t)=1\}.
\label{eq:service-completion}
\end{equation}
If the predicate never holds by the deadline, $\tau_s=\infty$. Each service earns its value only once, at first completion. A product's contribution must therefore be assessed along the dependency graph: its delivery may complete a service directly or make downstream work feasible.

\subsection{Product Processing and Output Timing}
Local processing can reduce the data to be transmitted, but occupies computing capacity and may change the quality of the delivered result. This tradeoff determines when a product becomes available for transmission and whether it can satisfy the service conditions above. Each ready product has a raw and a processed mode, described by their transmission volumes, computing demands, and quality outcomes. Fig.~\ref{fig:problem-model}(b) shows the two execution paths: raw data can enter the shared transmission queue immediately, whereas processed output must wait for computation to finish.

Product $i$ has arrival time $a_i$, deadline $d_i$, raw data size $D_i$, processing demand $P_i$, and processed-data ratio $\rho_i$. For selected mode $m_i$, its demands and quality are
\begin{equation}
\bigl(D_i(m_i),P_i(m_i),Q_i(m_i)\bigr)=
\begin{cases}
(D_i,0,1), & m_i=\mathrm{raw},\\
(\rho_iD_i,P_i,\widetilde Q_i), & m_i=\mathrm{process}.
\end{cases}
\label{eq:product-modes}
\end{equation}
The raw output enters the transmission queue as soon as its mode is selected. The processed output requires $P_i$ computing slots and becomes transmissible in the slot after computation finishes. A selected mode remains fixed until the product is delivered or expires. Only a fully delivered product enters $\mathcal R_t$; an incomplete product leaves execution at its deadline.

Raw delivery quality is normalized to one. The processed quality $\widetilde Q_i$ is drawn from a task-specific distribution with known support; its realization is revealed only after full delivery. The policy can use the distribution in planning and evaluate dependencies and acceptance predicates after observing the realized quality. The value of processing must therefore account for computing delay, reduced transmission volume, and the eventual quality decision; completing computation alone does not satisfy a service.

\subsection{Shared Computing and Dual-Access Resources}
Continuously arriving products contend for local computation and transmission. Cellular capacity is further limited by startup delay and usage budgets. Work performed for a current product can change which future services remain achievable, so computing, transmission, and cellular activation must be constrained jointly. We model one local processing unit and per-slot service capacity on each access path. Both links serve a shared transmission queue, while cellular startup state, traffic allowance, and activation allowance determine its usable supplementary capacity.

Let $x_{i,t}\in\{0,1\}$ indicate whether product $i$ receives computing service in slot $t$. Computing can advance only an arrived, ready product whose processed mode has been selected and is not yet complete. At most one product advances in a slot:
\begin{equation}
\sum_i x_{i,t}\leq 1.
\label{eq:compute-capacity}
\end{equation}
Let $S_t$ and $C_t$ be the Starlink and cellular transmission capacities in slot $t$, measured as transferable data volumes. Product $i$ receives actual volumes $b^S_{i,t}$ and $b^C_{i,t}$ on the two paths. Only queued outputs can use transmission capacity, and a product is delivered when its cumulative transmitted volume reaches the size of its selected mode. Cellular transmission is also bounded by remaining traffic allowance $B_t$, the policy's per-slot cap $\bar b_t$, and transmission-availability state $\zeta_t$:
\begin{equation}
\sum_i b^S_{i,t}\leq S_t,\qquad
\sum_i b^C_{i,t}\leq
\min\{\zeta_t C_t,\bar b_t,B_t\}.
\label{eq:access-capacity}
\end{equation}
$\zeta_t=1$ only when cellular startup has completed and the link can transmit in the current slot. Activating a previously disabled link incurs a startup wait, and startup slots consume activation allowance. Disabling the link resets startup. The traffic budget decreases with actual transmission, whereas the activation budget decreases in each active slot, even when no data can be sent. With $\ell_t$ indicating whether slot $t$ consumes an activation slot, total usage obeys
\begin{equation}
\sum_t\sum_i b^C_{i,t}\leq B_C,\qquad
\sum_t\ell_t\leq A_C,
\label{eq:cellular-budgets}
\end{equation}
where $B_C$ and $A_C$ are the episode's cellular traffic and activation budgets.

At each slot, the policy specifies an ordering for the shared queue. The executor serves that queue first with Starlink and then with any cellular capacity available in the same slot. Both paths may therefore advance the same product in one slot. Queue order determines which outputs receive priority, while activation timing must account for output availability and the remaining cellular allowance.

\subsection{Online Decisions and Objective}
\label{sec:online-decisions}
The model couples service completion to product execution and shared resources, but access capacity and delivered quality remain uncertain when decisions are made. The policy must choose current work from available information while accounting for downstream effects. We represent scheduling as a mapping from the observed state to a joint action, followed by updates to product progress, remaining budgets, and delivery records. The objective aggregates the value of services completed on time.

The online observation $o_t$ contains arrived products and their progress, active services, past capacities on both links, revealed delivery qualities, remaining traffic and activation budgets, cellular state, and the known resource configuration. The policy may maintain memory $\mu_t$ of previous observations and decisions. It does not observe the realized capacity of the current or future slots or the quality realization of a product that has not yet been delivered. These become available after slot execution and full product delivery, respectively.

The joint action combines the four controls numbered in Fig.~\ref{fig:problem-model}(b): mode selection, computing priority, transmission order, and cellular activation with a transmission cap:
\begin{equation}
u_t=(\mathcal M_t,c_t,\sigma_t,e_t,\bar b_t),
\label{eq:joint-action}
\end{equation}
where $\mathcal M_t$ specifies mode choices for ready products whose modes have not yet been selected, $c_t$ is the product advanced by computing, $\sigma_t$ orders queued products, $e_t$ is the cellular activation command, and $\bar b_t$ caps cellular transmission. The common executor applies the timing and resource rules of the preceding subsections, adds newly delivered products to $\mathcal R_{t+1}$, checks service completion, and forms the next observation.

Let $\pi$ denote a scheduling policy, $\omega$ one realization of tasks, access capacities, and product qualities, and $\mathcal S(\omega)$ its set of services. The cumulative completed-service value is
\begin{equation}
V(\pi;\omega)=\sum_{s\in\mathcal S(\omega)}
w_s\,\mathbf 1\{\tau_s\leq d_s\}.
\label{eq:service-value}
\end{equation}
Under a given distribution of tasks, access conditions, and qualities, the joint scheduling objective is
\begin{equation}
\max_{\pi\in\Pi_{\mathrm{causal}}}
\mathbb E_{\omega}[V(\pi;\omega)],
\label{eq:scheduling-objective}
\end{equation}
where $\Pi_{\mathrm{causal}}$ contains policies that use only available observations and memory and satisfy product dependencies, mode persistence, processing and transmission timing, and resource limits. We assess candidate policies through complete-run evaluation on a development set. Policy code and selection configurations are fixed before comparison on the specified evaluation sets.

We assume historical service-capacity estimates for both paths are available. The evaluation prototype updates samples for both links at every step, including when cellular transmission is disabled; the measurement or probing cost of such observations would need calibration against a deployment interface. Historical capacities can inform prediction, and revealed qualities can inform dependency decisions.

The central challenge is the propagation of decisions through one service workflow: processing changes output timing and quality, delivery changes which downstream products remain needed, and current resource use changes future feasibility. SatCraft expresses these relationships in executable scheduling policies, uses execution feedback to revise their code, and matches retained policies to observed access conditions.

\section{SatCraft: Policy Optimization and Execution}
The joint scheduling problem in Section~III requires a policy to connect processing choices, quality-dependent task paths, and dual-access budgets under limited observation. SatCraft optimizes executable scheduling policies that encode these decisions. An LLM agent generates and revises policy code using execution feedback, and a context-dependent policy selector chooses among valid policies. This section defines the policies and selector, the feedback-driven optimization loop, and online joint scheduling with the deployed system.

\subsection{Framework and Optimization Decomposition}
\begin{figure}[!t]
  \centering
  \includegraphics[width=\linewidth]{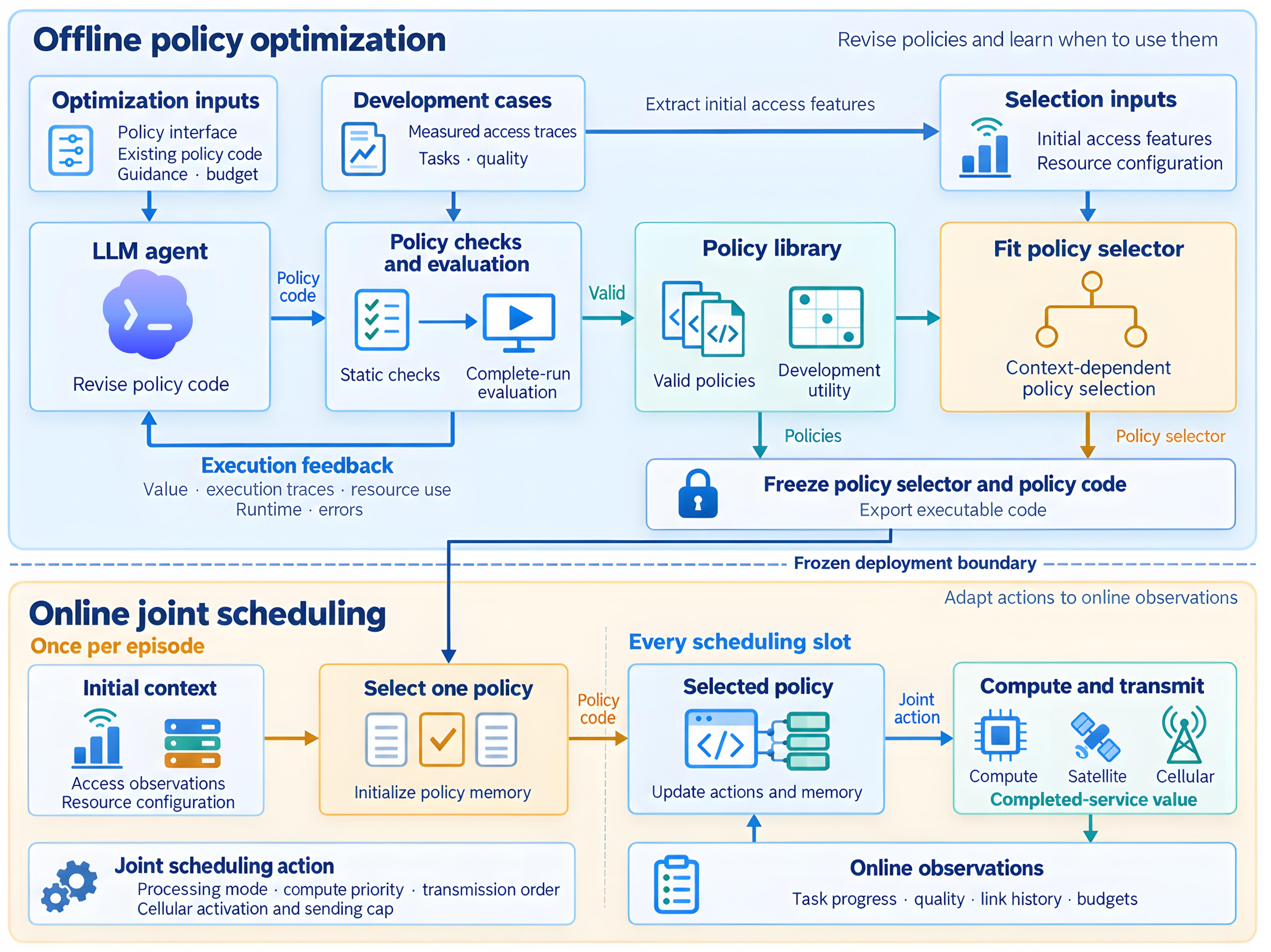}
  \caption{SatCraft's offline policy optimization and online joint scheduling.}
  \label{fig:satcraft-framework}
\end{figure}

Service completion depends on decisions whose effects span several slots. A policy must represent how processing, delivery, and contention from later tasks interact. Since the same logic may perform differently under different access conditions, useful policies must also be associated with the conditions under which they are deployed. SatCraft combines policy revision, which expands the available decision logic, with context-dependent policy selection, which determines where that logic is used. Fig.~\ref{fig:satcraft-framework} summarizes the data flow between offline policy optimization and online joint scheduling.

Each scheduling policy is implemented as executable code that maps observations and internal memory to joint scheduling actions. Let $c$ denote the policy code and $\pi_c$ the scheduling policy it implements. Given the observation and internal memory defined in Section~\ref{sec:online-decisions}, the policy returns a joint action and updates memory:
\begin{equation}
(u_t,\mu_{t+1})=\pi_c(o_t,\mu_t).
\label{eq:code-interface}
\end{equation}
All candidate policies use the same observation and action definitions, while their scheduling logic and memory updates may differ. The LLM agent may revise mode selection, resource anticipation, quality-dependent task handling, and cellular activation within this common interface.

Let $\mathcal P=\{\pi_j\}$ be the policy library containing candidate policies that pass the validity checks, and let $T_p$ be the policy selector for known resource configuration $p$. For measured access trace $r$, the policy selector maps initial access features $\phi_r$ to a policy index. The initial access observations and known resource configuration together define the context for policy selection. The development-set utility of the policy library combined with the selector is
\begin{equation}
\widehat J(\mathcal P,T)=\frac{1}{|\mathcal D_{\mathrm{dev}}|}
\sum_{\omega_{r,p,k}\in\mathcal D_{\mathrm{dev}}}
V\!\left(\pi_{T_p(\phi_r)};\omega_{r,p,k}\right),
\label{eq:portfolio-utility}
\end{equation}
where $k$ indexes quality realizations and $V$ is the completed-service value in~\eqref{eq:service-value}. Policy revision adds valid candidate policies to $\mathcal P$; selection fits $T=\{T_p\}$ from their development utilities. Code search proceeds through budgeted interactions, whereas the selector is chosen from a finite set of shallow tree structures.

After offline policy optimization, the policy selector and the code of its referenced policies are frozen. The policy selector uses initial access observations and the known resource configuration to select one policy per episode; the selected policy subsequently updates its actions and memory as tasks and resources evolve.

\subsection{Execution-Feedback-Driven Policy Optimization}
\label{sec:program-optimization}
A local change in an action can alter output availability, link allowance, and later service completion. Policy revisions must therefore undergo complete-run evaluation. Runtime errors and timeouts can also invalidate a candidate policy even when some local decisions appear beneficial. We organize code editing as a budgeted, multi-turn interaction in which the LLM agent receives the current policy code, case outcomes, and execution traces. Policy checks and evaluation combine static checks with complete-run evaluation. Valid policies enter the policy library, and execution feedback informs subsequent revisions.

For repair branch $b$, a researcher supplies interface specification $\mathcal I$, repair direction $d_b$, and budget $\mathcal B_b$, starting from an existing policy. Let $\bar c_{b,k}$ be the working policy code before submission $k$ and $\mathcal H_{b,k}$ the available code, tool outputs, and execution feedback. Submission of candidate policy code is represented by
\begin{equation}
c_{b,k}=\mathcal G_{\mathrm{LLM}}
\!\left(\mathcal I,d_b,\bar c_{b,k},\mathcal H_{b,k};\mathcal B_b\right).
\label{eq:agent-proposal}
\end{equation}
$\mathcal G_{\mathrm{LLM}}$ denotes the interaction that produces one submitted version of the policy code; a submission can involve several model responses and tool calls. The interface describes observation fields, joint actions, memory updates, and execution constraints. The repair direction focuses attention on an observed scheduling failure.

After submission, the source snapshot is checked for permitted imports, runtime file access, and dynamic evaluation. A failing check returns an error without complete-run evaluation. A policy that passes these checks is executed on the shared development cases. For case $n$, we record validity $\nu_n$, completed value $V_n$, resource use $\mathbf r_n$, runtime cost $\mathbf q_n$, error $\epsilon_n$, and execution trace $\Gamma_n$:
\begin{equation}
\mathcal F(c)=
\left\{(\nu_n,V_n,\mathbf r_n,\mathbf q_n,
\epsilon_n,\Gamma_n)\right\}_{n=1}^{|\mathcal D_{\mathrm{dev}}|}.
\label{eq:program-feedback}
\end{equation}
An execution trace records available observations, policy actions, actual resource use, and subsequent deliveries. These records provide execution feedback for offline policy revision; online observations provide the information used by a policy to choose its next action. Case validity requires actions to satisfy dependency and mode constraints and policy execution to finish within callback and episode time limits. An execution that violates these constraints has $\nu_n=0$. With $\operatorname{Static}(c)$ denoting the static-check result, a candidate policy is admissible when
\begin{equation}
A(c)=\mathbf 1\{\operatorname{Static}(c)=1\}
\prod_{n=1}^{|\mathcal D_{\mathrm{dev}}|}\nu_n.
\label{eq:admissibility}
\end{equation}
Starting from an existing policy library $\mathcal P_0$, the updated policy library is
\begin{equation}
\mathcal P=\mathcal P_0\cup
\left\{\pi_{c_{b,k}}:A(c_{b,k})=1\right\}.
\label{eq:candidate-library}
\end{equation}
Admissibility determines retention; utility under particular conditions determines deployment in the next subsection. A candidate policy with lower aggregate value may still be useful in some access conditions. All submissions and failures remain in the policy optimization record.

In the current policy optimization round, one repair estimates future contention from arrived tasks and their observed batch cadence, incorporating anticipated demand into computing, transmission, and cellular-budget projections. Another repair jointly considers per-slot supply, startup delay, remaining activation allowance, and computing-output timing to revise when supplementary access is activated. Delivered quality and the support of undelivered quality outcomes inform downstream task paths. These structural changes are implemented in different candidate policies and assessed through the same complete-run evaluation interface.

\subsection{Context-Dependent Policy Selection}
Candidate policies may make different tradeoffs among computing delay, transmission opportunities, and cellular allowance under different access conditions. To exploit these differences, policy utility must be associated with information available when a policy is selected. For each known resource configuration, SatCraft fits a shallow tree using initial access observations. Each leaf selects a policy from the library according to development-set utility.

The selector uses access observations from slots $0$--$7$; the first task batch arrives at slot $8$. Its 11-dimensional feature vector $\phi_r$ includes mean capacities of both links, the cellular-to-satellite mean ratio, coefficients of variation, trends, zero-capacity fractions, and the ratios of final to mean capacity. For each measured access trace $r$ and configuration $p$, utility is averaged over quality realizations:
\begin{equation}
U_{r,p}(\pi_j)=\frac{1}{K}
\sum_{k=1}^{K}V(\pi_j;\omega_{r,p,k}),
\label{eq:trace-utility}
\end{equation}
where $K$ is the number of development-set quality realizations for that trace--configuration pair.

For leaf $L$ containing a set of development traces, the selected policy index and leaf score are
\begin{equation}
\begin{split}
j^\star(L)&=\arg\max_{j:\pi_j\in\mathcal P}
\sum_{r\in L}U_{r,p}(\pi_j),\\
Q(L)&=\sum_{r\in L}U_{r,p}(\pi_{j^\star(L)}).
\end{split}
\label{eq:leaf-selection}
\end{equation}
For feature $f$ and threshold $\eta$, a split into $L_{\mathrm{left}}$ and $L_{\mathrm{right}}$ has gain
\begin{equation}
\Delta(L,f,\eta)=Q(L_{\mathrm{left}})+Q(L_{\mathrm{right}})-Q(L).
\label{eq:split-gain}
\end{equation}
The fitter considers thresholds between adjacent observed feature values and recursively takes the largest positive gain satisfying leaf-size constraints. Maximum tree depth is three, and each leaf contains at least three development traces.

SatCraft also evaluates three common preset tree structures. It keeps their features and thresholds but relabels their leaves from the current policy library. Let $\mathcal C_p$ contain the newly fitted tree and these three relabeled trees. The final policy selector is
\begin{equation}
T_p\in\arg\max_{T'\in\mathcal C_p}
\sum_r U_{r,p}\!\left(\pi_{T'(\phi_r)}\right).
\label{eq:selector-choice}
\end{equation}
The selector is chosen from four candidate trees according to the development utility of the policies assigned to their leaves. This selection step completes the offline workflow in Algorithm~\ref{alg:satcraft-offline}: feedback-driven repair expands the valid policy library, and selector fitting associates these policies with initial access observations for each resource configuration. The selectors and their referenced policy code are then exported and frozen. Complete-run evaluation of this combined system checks policy selection and execution before deployment for online joint scheduling.

\begin{algorithm}[!t]
\caption{Offline Policy Optimization and Selection}
\label{alg:satcraft-offline}
\begingroup
\renewcommand{\algorithmicrequire}{\textbf{Input:}}
\renewcommand{\algorithmicensure}{\textbf{Output:}}
\begin{algorithmic}[1]
\REQUIRE Interface $\mathcal I$, development cases $\mathcal D_{\mathrm{dev}}$, policy library $\mathcal P_0$, repair directions $d_b$ and budgets $\mathcal B_b$, feature map $\Phi$, tree constraints and preset structures.
\STATE Initialize $\mathcal P\leftarrow\mathcal P_0$ with existing policy utilities.
\FOR{each repair branch $b$}
  \WHILE{another submission is permitted by $\mathcal B_b$}
    \STATE Use the LLM agent to revise the working policy code and submit candidate $c$.
    \STATE Snapshot and statically check $c$; if checks pass, evaluate complete runs on $\mathcal D_{\mathrm{dev}}$ and record $\mathcal F(c)$.
    \IF{$A(c)=1$}
      \STATE $\mathcal P\leftarrow\mathcal P\cup\{\pi_c\}$.
    \ENDIF
    \STATE Return check errors or execution feedback to the LLM agent for subsequent revisions.
  \ENDWHILE
\ENDFOR
\STATE Extract initial access features $\phi_r$ using $\Phi$ for each development trace $r$.
\STATE Aggregate quality realizations into $U_{r,p}(\pi_j)$ for each trace and configuration.
\STATE Fit shallow trees, relabel preset trees, and choose each $T_p$ by development utility.
\STATE Export and freeze the policy selectors and referenced policy code; evaluate the exported system over complete runs.
\ENSURE Frozen policy selectors $T=\{T_p\}$ and the code of their referenced policies.
\end{algorithmic}
\endgroup
\end{algorithm}

\subsection{Policy Deployment and Online Joint Scheduling}
Online joint scheduling requires responses to changing task progress, revealed quality, and remaining budgets. SatCraft deploys the policy selector together with the code of the policies it references. After observing the initial access period, the selector chooses one policy for the episode using those observations and the known resource configuration. The selected policy then updates its actions and memory each slot; its code and the selector remain fixed after offline policy optimization.

For known configuration $p$ and initial features $\phi_r$,
\begin{equation}
j=T_p(\phi_r),\qquad
(u_t,\mu_{t+1})=\pi_j(o_t,\mu_t),\quad t\geq t_0.
\label{eq:frozen-execution}
\end{equation}
The index $j$ remains fixed within an episode, while the selected policy adapts its actions and memory to arrivals, dependencies, delivered quality, cellular state, and budgets. At each slot, the executor applies processing-mode choices, computing priority, transmission order, and cellular controls. The resulting task progress, budget updates, and delivery outcomes form the next observation, completing the execution loop in Algorithm~\ref{alg:satcraft-online}. Initial selection matches a policy to access conditions, and its subsequent decisions respond to changing service and resource states. This allows SatCraft to reuse scheduling logic refined from service failures while coordinating computation and transmission throughout service delivery.

\begin{algorithm}[!t]
\caption{Online Joint Computation and Transmission Scheduling}
\label{alg:satcraft-online}
\begingroup
\renewcommand{\algorithmicrequire}{\textbf{Input:}}
\renewcommand{\algorithmicensure}{\textbf{Output:}}
\begin{algorithmic}[1]
\REQUIRE Frozen policy selectors $T=\{T_p\}$ and their referenced policy code, feature map $\Phi$, known resource configuration $p$, and online observations.
\STATE Observe the initial access period and compute $\phi_r$ using $\Phi$.
\STATE Select $j\leftarrow T_p(\phi_r)$ for the episode.
\STATE Initialize policy memory $\mu_{t_0}$.
\FOR{each scheduling slot $t\geq t_0$}
  \STATE Read the current observation $o_t$.
  \STATE Compute $(u_t,\mu_{t+1})\leftarrow\pi_j(o_t,\mu_t)$.
  \STATE Execute computing and transmission according to $u_t$.
  \STATE Update product progress, budgets, delivery records, and completed services.
\ENDFOR
\ENSURE A joint scheduling action $u_t$ at each scheduling slot.
\end{algorithmic}
\endgroup
\end{algorithm}

\section{Performance Evaluation}

\subsection{Experimental Setup}
\subsubsection{Measured Access and Service Workloads}
We use the Alaska and Hawaii portions of the NUWiNS satellite--cellular drive measurements, collected in June and August 2024, respectively~\cite{wang2025digitaldivide}. Each selected trace pairs Transmission Control Protocol (TCP) downlink throughput measurements from Starlink and AT\&T or Verizon collected during the same drive. These paired rate sequences provide exogenous access-capacity inputs to the joint computation-and-transmission scheduling prototype.

We average adjacent measurement samples into nominal 1-s replay slots and retain the first 119 slots of each selected segment, preserving run boundaries and splitting at timestamp gaps longer than 1~s. Eligible segments have an average Starlink rate of at least 1~Mbit/s during the first eight slots. Table~\ref{tab:trace-characteristics} summarizes the 60 distinct selected traces. Rate percentiles and zero-rate fractions pool the 3,570 replay slots within each region, counting each selected trace once across repeated resource profiles and quality realizations. The data expose scheduling policies to broad throughput ranges and zero-throughput slots on both access paths.

\begin{table}[!t]
\centering
\caption{Characteristics of the selected access traces.}
\label{tab:trace-characteristics}
\begingroup
\small
\renewcommand{\arraystretch}{1.16}
\setlength{\tabcolsep}{3pt}
\begin{tabular}{@{}>{\raggedright\arraybackslash}m{0.46\columnwidth}
  *{2}{>{\raggedleft\arraybackslash}m{\dimexpr0.27\columnwidth-2\tabcolsep\relax}}@{}}
\toprule
Property & Alaska & Hawaii \\
\midrule
Collection (2024) & Jun.~21--26 & Aug.~16--19 \\
Paired traces & 30 & 30 \\
AT\&T traces & 19 & 14 \\
Verizon traces & 11 & 16 \\
Replay slot (s) & 1 & 1 \\
Slots per trace & 119 & 119 \\
\midrule
\multicolumn{3}{@{}l}{\textbf{Starlink throughput}} \\[2pt]
Median (Mbit/s) & 69.18 & 91.73 \\
P5--P95 (Mbit/s) & 2.10--155.76 & 0.00--153.63 \\
Zero-rate slots (\%) & 4.51 & 8.04 \\
\midrule
\multicolumn{3}{@{}l}{\textbf{Cellular throughput}} \\[2pt]
Median (Mbit/s) & 7.87 & 8.91 \\
P5--P95 (Mbit/s) & 0.00--147.84 & 0.00--87.33 \\
Zero-rate slots (\%) & 17.51 & 21.15 \\
\bottomrule
\end{tabular}
\endgroup
\end{table}

To evaluate service delivery under these access conditions, we pair the measured traces with modeled workloads that include task dependencies and quality requirements. Each case has six overlapping task batches, 24 product tasks, and 12 services. Service values are four or eight, giving a maximum completed value of 72 per case. The tasks share one local processor, and each product can be sent raw or processed locally before its smaller output is transmitted. Processing costs and result qualities are generated by the workload model. The realized quality of a processed result is revealed only after full delivery and is then used to evaluate downstream task dependencies and service-acceptance conditions.

Each case spans 119 slots. The first eight provide access observations; task batches arrive at slots 8, 24, 40, 56, 72, and 88. All methods face identical tasks, link sequences, quality realizations, and computing, transmission, and cellular-resource constraints. Table~\ref{tab:setup} summarizes the data splits and resource profiles; Quality denotes the number of quality seeds per trace--profile pair. Demand multipliers are relative to calibrated access capacity, and the computing multiplier scales product processing demand. Cellular startup and activation allowances are measured in slots.

\begin{table}[!t]
\centering
\caption{Data splits and resource profiles.}
\label{tab:setup}
\begingroup
\footnotesize
\renewcommand{\arraystretch}{1.16}
\setlength{\tabcolsep}{2.5pt}
\begin{tabular}{@{}>{\raggedright\arraybackslash}m{0.34\columnwidth}
  *{4}{>{\centering\arraybackslash}m{\dimexpr0.165\columnwidth-2\tabcolsep\relax}}@{}}
\toprule
\multicolumn{5}{@{}l}{\textbf{Data splits}} \\[3pt]
Split & Traces & Profiles & Quality & Cases \\
\midrule
Development set & 30 & 4 & 3 & 360 \\
Validation set & 30 & 4 & 2 & 240 \\
Test set & 30 & 4 & 3 & 360 \\
\addlinespace[5pt]
\multicolumn{5}{@{}l}{\textbf{Resource profiles}} \\[3pt]
Profile & \shortstack{Demand\\multiplier} & \shortstack{Compute\\multiplier}
        & \shortstack{Startup\\(slots)} & \shortstack{Activation\\(slots)} \\
\midrule
Balanced & 1.25 & 1.0 & 2 & 60 \\
Network pressure & 1.75 & 1.0 & 2 & 60 \\
Computing pressure & 1.25 & 1.4 & 2 & 60 \\
Cellular-activation pressure & 1.25 & 1.0 & 4 & 45 \\
\bottomrule
\end{tabular}
\endgroup
\end{table}

The development set is used to train the learning baselines and select their checkpoints, and to generate and repair SatCraft policies, tune parameters, and fit the policy selector. The validation set evaluates SatCraft and its ablation variants under different realizations of processed-result quality while retaining the development access traces and workload configuration. The test set contains 30 different measured access traces and supports the overall performance comparison, resource-pressure analysis, and online-cost evaluation. Together, the three sets comprise 960 cases derived from 60 measured runs.

\subsubsection{Compared Scheduling Methods}
We compare SatCraft with five baselines covering deadline priorities, rolling planning, resource-price optimization, and reinforcement learning. The three published edge-computing methods are adapted to the common service model, joint-action interface, and completed-service-value objective. The five baselines are:

\begin{itemize}
\setlength{\itemsep}{2pt}
\setlength{\parsep}{0pt}
\item \emph{Earliest-deadline-first (EDF)} prioritizes services by deadline and coordinates their prerequisite tasks, processing modes, and transmissions under quality and resource constraints. Its per-profile settings are selected on the development set from 216 heuristic configurations; all four selected configurations use deadline ordering.

\item \emph{Rolling} evaluates candidate service subsets and dispatch rules through receding-horizon simulation using recent access rates. It replans when services or delivery records change, or every six slots, and executes the first action of the selected plan. Development-set tuning selects four-slot rate-history windows for the balanced, network-pressure, and computing-pressure profiles and an eight-slot window for cellular-activation pressure.

\item \emph{Asynchronous deep progressive reinforcement learning (ADPRL)}~\cite{chen2024adprl} uses deep deterministic policy gradient learning with asynchronous experience collection, a single critic, target networks, and uniform replay. Its actor produces processing choices, computing and transmission priorities, and cellular controls. The adapted critic uses features of the resulting categorical actions and transmission order, with differentiable approximations for training.

\item \emph{EdgeOPT}~\cite{yang2024edgeopt} combines dynamic resource pricing, service admission, and minimum-rate water-filling. The adaptation maps resource shares to single-processor scheduling and transmission order, and checks complete-service feasibility under predicted rates and possible quality branches. Six price/history configurations are evaluated on 32 development cases; the selected price scale is 0.01 and the rate-history window is four slots.

\item \emph{Transformer-enhanced distributed deep reinforcement learning (TF-DDRL)}~\cite{wang2025tfddrl} uses a distributed actor--learner architecture with Gated Transformer-XL, V-trace correction, and prioritized trajectory replay. Its policy maps task and access history to the same joint scheduling actions. The network has two attention layers, four heads per layer, and a 16-step memory.
\end{itemize}

\subsubsection{Implementation Details}
ADPRL and TF-DDRL are trained on the development set for approximately 1.5 million transitions per seed. We use three independent seeds and select each seed's checkpoint by development-set completed-service value. Their rewards combine incremental completed-service value with shaping based on observable task progress.

SatCraft builds a policy library through code generation and feedback-driven repair, then selects a policy from initial access observations. For the current generation and repair round, the LLM agents run in OpenAI Codex with \texttt{gpt-6.1-sol} at xhigh reasoning effort. Each repair branch allows up to six policy submissions and 12 diagnostic calls. The library comprises 64 policies: 32 fixed-structure parameter policies, 23 previously generated policies, and nine valid candidates from the current repair round, including one with future-load anticipation disabled. The fitted selectors collectively reference 20 policies. Each resource profile has a decision tree of depth at most three, with at least three development traces per leaf; selection uses 11 features from the first eight slots. A selected policy runs for the whole episode and updates its actions and memory from online task and resource observations.

All methods use arrived tasks, delivery records, historical access rates, and known resource configurations as scheduling inputs and execute under the same physical constraints. The reported complete test runs use a 2-s callback limit and a 90-s case limit for every method. Policy code, model weights, and selection configurations remain fixed during evaluation.

We quantify offline cost using policy evaluations, model responses, and noncached input-plus-output tokens. Online costs are the time per scheduling callback and the total simulation time per case. Neural training uses an NVIDIA A100 GPU, and online measurements use an Intel Xeon Platinum 8358P CPU. Training time is reported as the mean and standard deviation.

\subsubsection{Metrics and Statistical Procedure}
The primary metric is cumulative completed-service value, computed as $\sum_{i=1}^{N}V_{A,i}$ over $N$ evaluation cases. Here, $V_{A,i}$ is the value achieved by method $A$ on case $i$, as defined in~\eqref{eq:service-value}. We also report completed services and fully served cases; a fully served case completes all 12 services by their deadlines with acceptable quality. The relative gain of method $A$ over $B$ is
\begin{equation}
G_{A,B}=\left(\frac{\sum_{i=1}^{N}V_{A,i}}{\sum_{i=1}^{N}V_{B,i}}-1\right)\times100\%.
\label{eq:relative-gain}
\end{equation}
Performance results for ADPRL and TF-DDRL are averaged over three training seeds. We compute 95\% confidence intervals using 20,000 paired bootstrap resamples clustered by measured trace.

\subsection{Overall Service-Completion Performance}
\begin{figure}[!t]
\centering
\includegraphics[width=\linewidth]{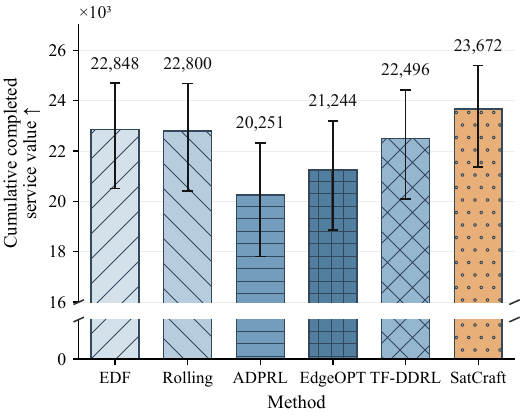}
\caption{Cumulative completed-service value on the test set.}
\label{fig:overall}
\end{figure}

We evaluate whether SatCraft improves both the value of delivered services and the fulfillment of service requests. We compare SatCraft with EDF, Rolling, ADPRL, EdgeOPT, and TF-DDRL. All six methods execute the same 360 test cases with identical link sequences, service arrivals, quality realizations, and physical resource budgets. Fig.~\ref{fig:overall} compares cumulative completed-service value, and Table~\ref{tab:service-completion} reports completed services and fully served cases.

SatCraft achieves the highest cumulative value, 23,672, exceeding the five baselines by 3.61\%--16.89\%. Its gains over TF-DDRL, ADPRL, and EdgeOPT are 5.23\%, 16.89\%, and 11.43\%, respectively. These improvements show that SatCraft's optimized scheduling policies deliver greater service value under the same computing and access budgets.

\begin{table}[!t]
\centering
\caption{Service completion on the test set.}
\label{tab:service-completion}
\begingroup
\small
\renewcommand{\arraystretch}{1.2}
\setlength{\tabcolsep}{3pt}
\begin{tabular*}{\columnwidth}{@{\extracolsep{\fill}}lrr@{}}
\toprule
\multirow{2}{*}{Method} & Completed services & Fully served cases \\
 & {\footnotesize (of 4,320)} & {\footnotesize (of 360)} \\
\midrule
EDF & 3,832 & 223 \\
Rolling & 3,782 & 212 \\
ADPRL & 3,219 & 72 \\
EdgeOPT & 3,387 & 111 \\
TF-DDRL & 3,745 & 199 \\
\addlinespace[1pt]
\textbf{SatCraft} & \textbf{3,935} & \textbf{270} \\
\bottomrule
\end{tabular*}
\endgroup
\end{table}

Table~\ref{tab:service-completion} shows that the value gain is accompanied by more completed services and fully served cases. SatCraft completes 3,935 of the 4,320 requested services, or 91.09\%, and fully serves 270 of the 360 cases, or 75.0\%. The baselines fully serve between 72 and 223 cases. A fully served case requires all 12 services to meet their deadlines and quality requirements, so the additional fully served cases reflect more complete delivery across competing requests within an episode.

Together, the higher completed value, service count, and fully served case count show that SatCraft improves end-to-end service fulfillment under shared computing and satellite--cellular access constraints.

\subsection{Performance Under Resource Pressure}
\begin{figure}[!t]
\centering
\includegraphics[width=\linewidth]{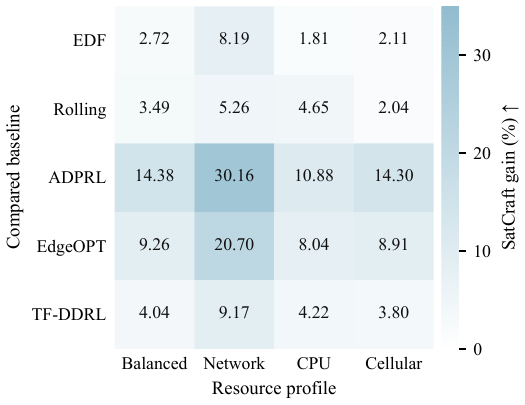}
\caption{SatCraft's relative completed-value gains across resource profiles.}
\label{fig:pressure}
\end{figure}

This experiment evaluates how SatCraft's service-delivery performance varies across computing, transmission, and cellular-activation constraints. We partition the 360 test cases into the four profiles in Table~\ref{tab:setup}, giving 90 cases per profile. The comparison uses the same six methods and averaging procedure as the overall experiment, with cumulative values displayed as integers. Within each profile, all methods receive identical measured traces, tasks, quality outcomes, and physical resource budgets. The labels Balanced, Network, CPU, and Cellular denote the balanced, network-pressure, computing-pressure, and cellular-activation-pressure profiles, respectively. Fig.~\ref{fig:pressure} shows SatCraft's relative gains over the five baselines, and Table~\ref{tab:pressure} reports their cumulative completed-service values.

\begin{table}[!t]
\centering
\caption{Cumulative completed-service value under four resource profiles.}
\label{tab:pressure}
\begingroup
\small
\renewcommand{\arraystretch}{1.2}
\setlength{\tabcolsep}{3pt}
\begin{tabular*}{\columnwidth}{@{\extracolsep{\fill}}lrrrr@{}}
\toprule
Method & Balanced & Network & CPU & Cellular \\
\midrule
EDF & 5,884 & 5,324 & 5,752 & 5,888 \\
Rolling & 5,840 & 5,472 & 5,596 & 5,892 \\
ADPRL & 5,284 & 4,425 & 5,281 & 5,260 \\
EdgeOPT & 5,532 & 4,772 & 5,420 & 5,520 \\
TF-DDRL & 5,809 & 5,276 & 5,619 & 5,792 \\
\addlinespace[1pt]
\textbf{SatCraft} & \textbf{6,044} & \textbf{5,760} & \textbf{5,856} & \textbf{6,012} \\
\bottomrule
\end{tabular*}
\endgroup
\end{table}

SatCraft achieves the highest cumulative completed-service value in every resource profile, and all 20 relative gains in Fig.~\ref{fig:pressure} are positive. Against ADPRL, EdgeOPT, and TF-DDRL, its gains range from 3.80\% to 30.16\% across the four profiles. The advantage therefore persists as the limiting resource shifts between computation, transmission, and cellular activation.

The largest gain over each of these three methods occurs under network pressure: 30.16\% over ADPRL, 20.70\% over EdgeOPT, and 9.17\% over TF-DDRL. This pattern shows that SatCraft's optimized scheduling policies maintain their delivery advantage across the evaluated resource bottlenecks, with particularly large improvements when greater transmission demand intensifies competition for available link capacity.

\subsection{Ablation of the Policy Library and Context-Dependent Selection}
We evaluate SatCraft and five ablation variants on the validation set to examine the contributions of generated policies, current repairs, and context-dependent policy selection to completed-service value. When candidate policies are removed, the selector is refitted on the development set under the same features and tree constraints.

The new-repair ablation removes all current future-load and link-repair policies. The future-load and new-link ablations remove only the corresponding policies; the future-load ablation retains the matched disabled version. The generated-structure ablation retains only parameter policies with the same selector, while the selection ablation fixes one policy per resource profile from the full library. Table~\ref{tab:ablation} reports completed values and SatCraft's gain over each variant.

\begin{table}[!t]
\centering
\caption{Ablation results on the validation set.}
\label{tab:ablation}
\begingroup
\footnotesize
\renewcommand{\arraystretch}{1.18}
\setlength{\tabcolsep}{3pt}
\begin{tabular}{@{}>{\raggedright\arraybackslash}m{0.56\columnwidth}
  *{2}{>{\raggedleft\arraybackslash}m{\dimexpr0.22\columnwidth-2\tabcolsep\relax}}@{}}
\toprule
Configuration & \shortstack[r]{Completed\\value $\uparrow$} & \shortstack[r]{SatCraft\\gain} \\
\midrule
Without new repairs & 14,688 & +0.71\% \\
Without future-load repair & 14,760 & +0.22\% \\
Without new link repair & 14,748 & +0.30\% \\
Without generated structures & 14,556 & +1.62\% \\
Without context-dependent selection & 14,516 & +1.90\% \\
\midrule
\textbf{SatCraft} & \textbf{14,792} & -- \\
\bottomrule
\end{tabular}
\endgroup
\end{table}

SatCraft's completed value of 14,792 exceeds that of all five ablations. Restricting the library to parameter policies yields 14,556; the full library adds 236, a 1.62\% increase. Fixing one policy per profile from the full library yields 14,516; context-dependent policy selection adds 276, or 1.90\%. These comparisons support the contribution of generated policy structures and context-dependent policy selection when processed-result qualities change on the same access traces.

Removing all current repairs reduces value by 104, corresponding to a 0.71\% SatCraft gain. Removing future-load or new link-repair policies separately reduces value by 32 or 44, corresponding to 0.22\% and 0.30\%. Thus, the current repair round adds a smaller incremental benefit. Because the selector is refitted after policy removal, these differences measure changes in the library and its fitted selection jointly. Overall, the validation results support the role of a varied policy library and context-dependent policy selection, with a further contribution from the current repairs.

\subsection{Policy Optimization Cost and Runtime Overhead}
\label{sec:construction-cost}
We evaluate the offline optimization effort and online execution cost associated with SatCraft's service-delivery gains, using EDF, ADPRL, and TF-DDRL as representative heuristic and learning-based baselines.

\begin{figure}[!t]
\centering
\includegraphics[width=\linewidth]{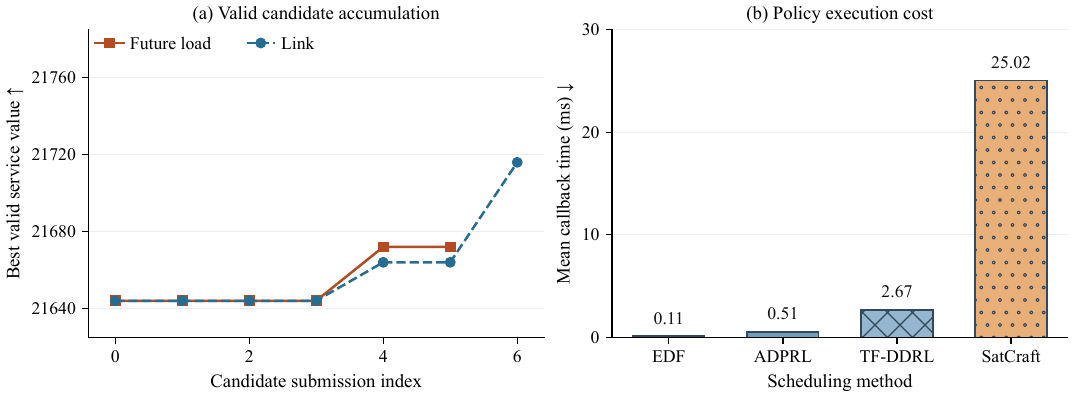}
\caption{Policy-score progression during repair and online scheduling cost.}
\label{fig:cost}
\end{figure}

The offline analysis covers SatCraft's current repair round: two parallel branches for future-load anticipation and link activation, starting from a policy with a development value of 21,644. We also report the training costs of ADPRL and TF-DDRL. We measure online costs on the test set under the common timing protocol. Fig.~\ref{fig:cost} shows repair progress and mean callback time, while Tables~\ref{tab:repair-cost} and~\ref{tab:runtime} report offline and online costs, respectively.

\begin{table}[!t]
\centering
\caption{Offline optimization cost and outcomes.}
\label{tab:repair-cost}
\begingroup
\footnotesize
\renewcommand{\arraystretch}{1.18}
\setlength{\tabcolsep}{3pt}
\begin{tabular}{@{}>{\raggedright\arraybackslash}m{0.40\columnwidth}
  *{3}{>{\raggedleft\arraybackslash}m{\dimexpr0.20\columnwidth-2\tabcolsep\relax}}@{}}
\toprule
\multicolumn{4}{@{}l}{\textbf{SatCraft: current repair round}} \\[3pt]
Metric & \shortstack[r]{Future-load\\repair} & \shortstack[r]{Link\\repair} & Total \\
\midrule
Policy evaluations & 5 & 6 & 11 \\
Model responses & 42 & 37 & 79 \\
Tokens & 102,633 & 115,720 & 218,353 \\
Record window (min) & 16.73 & 14.81 & -- \\
\addlinespace[3pt]
Valid policies & 3/5 & 6/6 & 9/11 \\
Best valid value $\uparrow$ & 21,672 & 21,716 & -- \\
\midrule
\multicolumn{4}{@{}l}{\textbf{Learning baselines: training per seed}} \\[3pt]
Method & \shortstack[r]{Time\\(min)} & \shortstack[r]{Sampled\\transitions} & \shortstack[r]{Development\\cases} \\
\midrule
ADPRL & $9.98\pm0.16$ & 1,500,032 & 3,960 \\
TF-DDRL & $16.96\pm0.70$ & 1,500,032 & 2,160 \\
\bottomrule
\end{tabular}
\endgroup
\end{table}

Across both SatCraft branches, 11 candidate policies were evaluated, nine passed validity checks, and the process used 79 model responses and 218,353 noncached input-plus-output tokens. The best future-load and link-repair scores were 21,672 and 21,716, increases of 28 and 72 over the common starting point. The parallel branches had record windows of 16.73 and 14.81~min. Fig.~\ref{fig:cost}(a) shows how valid improvements accumulate as candidates are evaluated. Two future-load policies were rejected for timeout; a subsequent bounded-search policy was valid on all 360 development cases. Execution feedback therefore supports policy improvement while checking that the added logic remains executable within the prescribed budget.

Table~\ref{tab:repair-cost} also reports training times of $9.98\pm0.16$ and $16.96\pm0.70$~min per seed for ADPRL and TF-DDRL, respectively. These training runs produce the neural policies used for the online comparison; the SatCraft records quantify feedback-driven repair of its existing policy code.

\begin{table}[!t]
\centering
\caption{Online scheduling cost on the test set.}
\label{tab:runtime}
\begingroup
\small
\renewcommand{\arraystretch}{1.25}
\setlength{\tabcolsep}{3pt}
\begin{tabular*}{\columnwidth}{@{\extracolsep{\fill}}lrrrr@{}}
\toprule
\multirow{2}{*}{Method} & \multicolumn{2}{c}{Callback time} & \multicolumn{2}{c@{}}{Case execution time} \\
\cmidrule(lr){2-3}\cmidrule(l){4-5}
 & Mean (ms) & Max (s) & Mean (s) & P95 (s) \\
\midrule
EDF & 0.11 & 0.003 & 0.021 & 0.025 \\
ADPRL & 0.51 & 0.006 & 0.071 & 0.074 \\
TF-DDRL & 2.67 & 0.080 & 0.328 & 0.333 \\
SatCraft & 25.02 & 1.180 & 2.986 & 8.320 \\
\bottomrule
\end{tabular*}
\endgroup
\end{table}

SatCraft's scheduling callbacks take 25.02~ms on average, and its longest callback takes 1.180~s, below the shared 2-s limit. EDF, ADPRL, and TF-DDRL have lower means of 0.11, 0.51, and 2.67~ms, respectively. SatCraft's additional runtime is consistent with the resource projections and cross-task coordination performed by its repaired policies. Together with the higher completed-service value in Fig.~\ref{fig:overall}, these timings show that SatCraft improves service delivery while meeting the evaluated online execution budget.

\subsection{Repair Mechanisms and Execution Traces}
We use paired policy variants and complete execution traces to examine how repairs change resource use and service completion, and how their effects vary across cases. Case A compares future-load anticipation enabled and disabled; case B compares the cellular-activation repair enabled and disabled. The first component infers a future-load template from arrived tasks and includes the next anticipated batch in joint projections of computing, transmission, and cellular budgets. The second includes per-slot supply, remaining activation allowance, and computed-output availability when deciding when cellular service is useful. 

Within each pair, only the relevant component switch changes; code elsewhere, tasks, measured links, and quality outcomes are held fixed. The cases use the same access trace and computing-pressure profile but different quality seeds. Each additional completed service is worth eight. Table~\ref{tab:paired-cases} gives final outcomes, and Table~\ref{tab:component-pairs} summarizes component effects on all 360 development cases; wins, ties, and losses compare enabled with disabled variants case by case. Both future-load variants were evaluated in the branch counted in Table~\ref{tab:repair-cost}; the cellular-activation and quality-path pairs use separate policy variants for verification. 

\begin{figure}[!t]
\centering
\includegraphics[width=\linewidth]{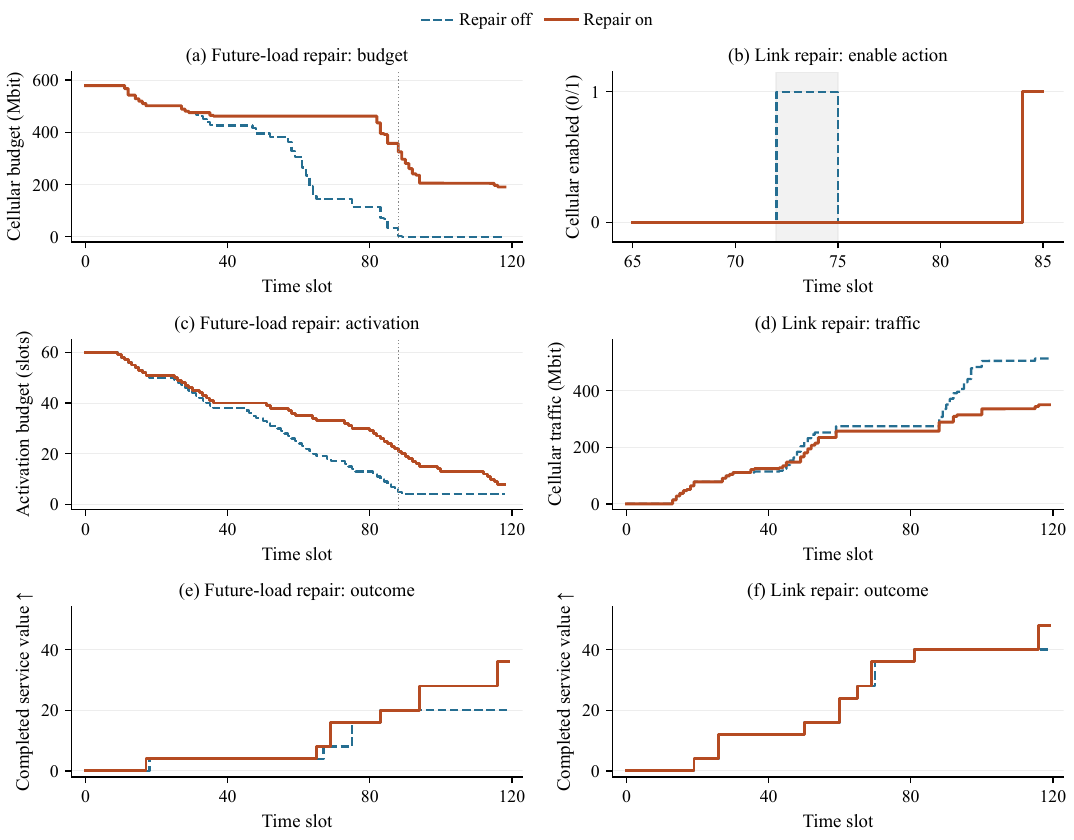}
\caption{Execution traces for paired repair cases.}
\label{fig:repair-cases}
\end{figure}

Fig.~\ref{fig:repair-cases} links resource states to completion. Its left panels (a), (c), and (e) show remaining cellular traffic, activation allowance, and cumulative completed value for case A. The right panels (b), (d), and (f) show activation actions, cumulative cellular traffic, and cumulative completed value for case B. Dashed and solid curves denote disabled and enabled variants, respectively; the vertical line at slot 88 marks a later batch arrival. Panel (b) focuses on slots 65--85, with slots 72--74 shaded; other panels cover the full case. Resource states are sampled before actions, whereas cumulative traffic and completed value are updated after execution.

\begin{table}[!t]
\centering
\caption{Resource use and service outcomes in two paired cases.}
\label{tab:paired-cases}
\begingroup
\small
\renewcommand{\arraystretch}{1.18}
\setlength{\tabcolsep}{3pt}
\begin{tabular}{@{}>{\raggedright\arraybackslash}m{0.60\columnwidth}
  *{2}{>{\raggedleft\arraybackslash}m{\dimexpr0.20\columnwidth-2\tabcolsep\relax}}@{}}
\toprule
Metric & Off & On \\
\midrule
\multicolumn{3}{@{}l}{\textbf{Case A: future-load anticipation}} \\[3pt]
Completed-service value & 20 & \textbf{36} \\
Cellular traffic (Mbit) & 578.83 & 388.21 \\
Activation slots used & 56 & 52 \\
Computing slots used & 57 & 54 \\
Traffic remaining at slot 88 (Mbit) & 2.28 & 325.21 \\
Activation slots remaining at slot 88 & 5 & 21 \\
Extra services completed (slots) & -- & 2 (94, 116) \\
\midrule
\multicolumn{3}{@{}l}{\textbf{Case B: cellular-activation repair}} \\[3pt]
Completed-service value & 40 & \textbf{48} \\
Cellular traffic (Mbit) & 513.30 & 350.16 \\
Activation slots used & 60 & 54 \\
Computing slots used & 43 & 43 \\
Activation actions at slots 72--74 & 3 & 0 \\
Cellular traffic at slots 72--74 (Mbit) & 0.00 & 0.00 \\
Extra services completed (slots) & -- & 1 (116) \\
\bottomrule
\end{tabular}
\endgroup
\end{table}

In case A, Figs.~\ref{fig:repair-cases}(a) and (c) show that at slot 88, after the later batch arrives but before that slot's action executes, the disabled policy has only 2.28~Mbit and five activation slots left, whereas the enabled policy retains 325.21~Mbit and 21 slots. Panel (e) and Table~\ref{tab:paired-cases} show that the enabled variant completes two additional services at slots 94 and 116, raising completed value from 20 to 36 while using less cellular traffic and computing time overall. Future-load anticipation reserves transmission and activation capacity for later work; the gain arises from allocation across batches rather than greater resource consumption.

In case B, Fig.~\ref{fig:repair-cases}(b) shows that the disabled variant activates cellular access in slots 72--74 without transmitting data; the enabled variant avoids those three activations and also changes subsequent transmission scheduling. Panels (d) and (f) and Table~\ref{tab:paired-cases} show cellular traffic falling from 513.30 to 350.16~Mbit and activation use from 60 to 54 slots, while computing use remains 43 slots. One additional service completes at slot 116, raising the value from 40 to 48. The repair reduces activation without useful transmission and changes later transfer timing, enabling an additional delivery with the same computing use.

\begin{table}[!t]
\centering
\caption{Paired component effects on the development set.}
\label{tab:component-pairs}
\begingroup
\small
\renewcommand{\arraystretch}{1.20}
\setlength{\tabcolsep}{3pt}
\begin{tabular}{@{}>{\raggedright\arraybackslash}m{0.31\columnwidth}
  *{3}{>{\raggedleft\arraybackslash}m{\dimexpr0.23\columnwidth-2\tabcolsep\relax}}@{}}
\toprule
Metric & \shortstack[r]{Future-load\\anticipation} & \shortstack[r]{Cellular\\activation} & \shortstack[r]{Quality-path\\handling} \\
\midrule
Off value & 21,432 & 21,644 & 21,732 \\
On value & 21,672 & 21,740 & 21,740 \\
Change (on--off) & +240 & +96 & +8 \\
\midrule
Improved cases & 71 & 19 & 7 \\
Tied cases & 240 & 337 & 347 \\
Degraded cases & 49 & 4 & 6 \\
\bottomrule
\end{tabular}
\endgroup
\end{table}

Table~\ref{tab:component-pairs} places these successful execution traces in the full development set. Future-load anticipation adds 240 in cumulative value but improves 71 cases and worsens 49. Cellular-activation repair adds 96, with 19 improvements and four regressions. Quality-path handling adds only eight; most cases are unchanged. The effects of code revisions therefore depend on case conditions, and successful examples do not represent every workload. Policy evaluation must account for net completed value and regressions across the full case batch. The two execution traces illustrate resource reservation across batches and the avoidance of ineffective cellular activation, while complete-run evaluation checks which revised policies satisfy the requirements for retention.

\section{Conclusion}
We studied joint computation and transmission scheduling for mobile edge services with satellite and cellular access. Our formulation combines processing modes, quality-dependent task paths, and cellular budgets in an objective based on timely completion of acceptable services. SatCraft combines LLM-guided policy optimization with context-dependent policy selection to turn execution feedback into reusable scheduling logic that coordinates processing and transmission for timely, acceptable service delivery. Experiments using Starlink and cellular measurement data show that SatCraft outperforms all baselines in cumulative completed-service value on the test set. Ablations on the validation set support the contribution of generated policy structures and context-dependent policy selection. Paired execution traces connect future resource reservation and cellular-activation changes to additional completed services. These results demonstrate the value of LLM agents for optimizing executable joint scheduling policies to improve service completion under the computing, communication, and quality constraints of satellite--cellular networks. Future work will calibrate processing costs and result quality for concrete inspection tasks, account for operational uplink access and measurement and device-energy costs, and test the policy library and selector in corresponding workflows.

\appendices
\section{Reproducible Prompt Descriptions for the LLM Agent}
This appendix presents two parameterized prompt templates for the policy optimization process in Section~\ref{sec:program-optimization}. The initial template specifies the task, policy interface, repair direction, and budgets; the continuation template incorporates execution feedback to guide subsequent policy revisions.

\subsection{Task Interface and Repair Direction}
The initial template combines the task, available context, policy interface, and repair direction. Italicized brackets indicate fields filled from the working policy code, development records, and branch configuration.

\begin{prompttemplate}{Initial policy optimization}
\footnotesize
\textbf{Task.} Improve the executable scheduling policy for satellite--cellular edge services. Maximize the cumulative value of services delivered by their deadlines with acceptable quality, subject to the service model and resource budgets.

\smallskip\noindent\textbf{Context.} \emph{[Service and product-task interface; quality dependencies; computing and wireless constraints; current policy code; development results and execution traces; repair direction; submission, diagnostic, and runtime budgets.]}

\smallskip\noindent\textbf{Policy interface.} Implement \texttt{policy(observation, memory)}. Return mode choices, a computing task, shared transmission-queue order, cellular activation, and a cellular sending cap; update memory if needed. Use only available observations and memory. Respect mode and timing constraints, and the static checks on imports, file access, and dynamic evaluation.

\smallskip\noindent\textbf{Revision.} Inspect the current logic and complete-run outcomes. Revise the policy code to address \emph{[repair direction]}, then submit it through the supplied tools. Use completed-service value, validity, resource states, runtime, and errors to guide bounded revisions.

\smallskip\noindent\textbf{Return.} Policy code, the reason for the change, and locations of complete evaluation and validity records.
\end{prompttemplate}

The two repair-direction fields in the current round are (i) \emph{future load and resource contention}: infer demand from arrived tasks and observed batch cadence to coordinate current work with later services' computing, transmission, and cellular needs; and (ii) \emph{cellular activation and output timing}: account for startup wait, activation budget, per-slot supply, and computed-output availability when assessing when supplementary access can transmit useful data. Interface modes \texttt{raw} and \texttt{compress} correspond to the raw and processed modes in the model. Each branch allows up to six submissions and 12 diagnostics; policy callback and episode limits are 2 and 90~s. 

\subsection{Continuation After Execution Feedback}
A continuation keeps the task and interface fixed while adding execution feedback for the next revision.

\begin{prompttemplate}{Continuation after execution feedback}
\footnotesize
\textbf{Task.} Continue improving \emph{[current policy code]} under the same interface, service objective, and repair direction.

\smallskip\noindent\textbf{Feedback.} \emph{[Value, validity, resource, and runtime summary; relevant cases and complete execution traces; errors and timeouts; remaining submission and diagnostic budget.]}

\smallskip\noindent\textbf{Diagnosis.} Identify which policy decisions affected output availability, resource use, and later service completion. Distinguish changes in individual actions from changes in service-completion outcomes.

\smallskip\noindent\textbf{Revision.} Modify the working policy code and submit the next candidate policy within the remaining budget. After errors or timeouts, reduce execution complexity. Keep the record of each submission and evaluation.

\smallskip\noindent\textbf{Return.} The revised policy code, reason for the revision, and corresponding complete-evaluation results.
\end{prompttemplate}

\bibliographystyle{IEEEtran}
\bibliography{ref}
\end{document}